\documentclass[]{aa501}

\usepackage{graphicx,amssymb,lscape,setspace,xspace}
\begin{document}
\titlerunning{Optical spectroscopy of the M\,15 IVC}
\title{Optical spectroscopy of the M\,15 intermediate velocity cloud
       \thanks{Based on observations made with the Apache Point Observatory 3.5-m, U.S.A.,
               Very Large Telescope, Chile, and William Herschel Telescope, La Palma, Spain.}
}


\author{J. V. Smoker
        \inst{1}
        \and
        F. P. Keenan
        \inst{1}
        \and
        N. Lehner 
        \inst{2}
        \and
        C. Trundle
        \inst{1,3}
        }       
\institute{Astrophysics and Planetary Science Division,
           Department of Pure and Applied Physics,
           The Queen's University of Belfast,
           University Road, Belfast, BT7 1NN, 
           U.K.
           \and 
           Department of Physics and Astronomy, 
           Johns Hopkins University, 
           3400 North Charles Street, Baltimore, 
           MD 21218, U.S.A
           \and
           Isaac Newton Group, Apartado de Correos 368, Santa Cruz de La Palma,
           Tenerife 38780, Canary Islands, Spain
	}
\offprints{j.smoker@qub.ac.uk}
\date{\bf{Received Accepted }}

\abstract{
We present echelle spectrograph observations in the Na\,D lines, at 
resolutions of 6.2$-$8.5 km\,s$^{-1}$, for 11 stars located in the line-of-sight to 
the M\,15 intermediate velocity cloud (IVC), which has 
a radial velocity of $\sim$ +70 km\,s$^{-1}$ in the Local Standard of Rest. This cloud is 
a part of IVC Complex gp. The targets range in magnitude from $m_{V}$=13.3$-$14.8. Seven of the 
observed stars are in the M\,15 globular cluster, the remaining four being field stars. 
Three of the observed cluster stars are located near a peak in intensity of the IVC H\,{\sc i} 
column density as observed at a resolution of $\sim$ 1 arcmin. Intermediate velocity gas 
is detected in absorption towards 7 stars, with equivalent widths in Na\,D$_{2}$ ranging from 
$\sim$ 0.09$-$0.20\AA, corresponding to log$_{10}$($N_{\rm Na}$ cm$^{-2}$) $\sim$ 11.8$-$12.5, and 
Na\,{\sc i}/H\,{\sc i} column density ratios (neglecting the H\,{\sc ii} component) ranging 
from $\sim$ (1$-$3)$\times$10$^{-8}$. 
Over scales ranging from 30 arcsec to 1 arcmin, the Na\,{\sc i} column density and the 
Na\,{\sc i}/H\,{\sc i} ratio varies by upto 70 per cent and a factor of $\sim$ 2, respectively. 
Combining the current sightlines with previously obtained Na\,{\sc i} data from Kennedy et al. (1998b), the 
Na\,{\sc i}/H\,{\sc i} column density ratio over cluster sightlines varies by upto a factor of 
$\sim$ 25, when using H\,{\sc i} 
data of resolution $\sim$2$\times$1 arcmin. One cluster star, M\,15 ZNG-1, 
was also observed in the Ca\,{\sc i} ($\lambda_{\rm air}$=4226.728\AA) and 
Ca\,{\sc ii} ($\lambda_{\rm air}$=3933.663\AA) lines. A column density ratio 
$N$(Ca\,{\sc i})/$N$(Ca\,{\sc ii}) $<$ 0.03 was found, typical of values seen in the 
warm ionised interstellar medium. Towards this sightline, the IVC has a 
Na\,{\sc i}/Ca\,{\sc ii} column density ratio of $\sim$ 0.25, similar to 
that observed in the local interstellar medium. Finally, we detect tentative evidence for IV absorption in 
K\,{\sc i} ($\lambda_{\rm air}$=7698.974\AA) towards 3 cluster stars, which have $N$(K\,{\sc i})/$N$(H\,{\sc i}) 
ratios of $\sim$ 0.5$-$3$\times$10$^{-9}$.
}
\maketitle

\section{Introduction}

Fine-scale structure in low velocity interstellar gas, over scales ranging from tens to 
hundreds of thousands of 
AU, has been found by many different workers using a variety of techniques, including observations of 
pulsars, globular clusters and binary stellar systems that act as tracers of the interstellar medium (e.g. Frail et al. 1994; 
Kennedy et al. 1998a; Andrews et al. 2001). The presence of this structure implies dense clumps of 
gas in otherwise diffuse sightlines, and large overpressures with respect to the interstellar thermal 
pressure of $\sim$ 3000 cm$^{-3}$ K (Welty \& Fitzpatrick 2001). Regardless of whether such 
clumps exist or are simply an illusion caused by geometric effects (Heiles 1997), the small-scale 
structures must be explained by any model of the interstellar medium. 

For gas at intermediate and high velocities, fewer studies of small-scale structure 
exist, due to the difficulty in finding relatively bright, closely-spaced probe objects 
at distances greater than the absorbing material. One obvious location for probe stars is 
globular clusters. Hence Kerr \& Knapp (1972) performed an absorption-line survey towards 
many such objects, in particular finding an intermediate-velocity cloud towards M\,15, a globular 
cluster that lies at a distance of $\sim$ 10 kpc (Carretta et 
al. 2000). This cloud is a part of the IVC Complex gp (Wakker 2001), a cloud 
complex located in the line-of-sight towards the HVC Complex GP. Parts  
of Complex gp are likely to lie at a distance exceeding 0.8 kpc (Little et al. 1994). 
The distance towards the M\,15 IVC is uncertain, although is likely to be less 
than 3kpc (Smoker et al. 2001). The study of parts of this cloud is the subject of the current paper. 

\begin{figure}
\includegraphics[]{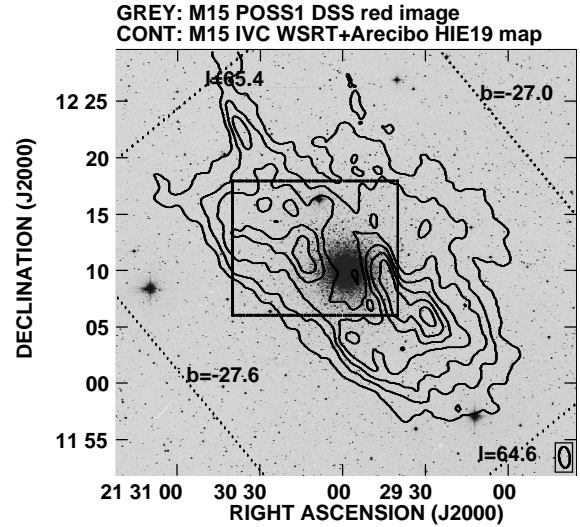}
\caption{WSRT plus Arecibo combined H\,{\sc i} column density map of the M\,15 
IVC integrated between +50 and +90 km\,s$^{-1}$ in the LSR and at a resolution of 
111$^{\prime\prime}\times$56$^{\prime\prime}$, overlaid on the 
red POSS-1 digital sky survey image towards M\,15 (Smoker et al. 2002). 
Contour levels are at $N_{\rm HI}$=(2,4,6,8,10,12,14)$\times$10$^{19}$ cm$^{-2}$.
Lines of constant Galactic longitude and latitude are also plotted on the 
figure. The rectangle shows the region depicted in Fig. \ref{samplefig}.
} 
\label{combhisur}
\end{figure}

Previous optical observations of the M\,15 IVC include low-resolution  
Ca\,{\sc ii} K ($\lambda_{\rm air}$=3933.663\AA) and Na\,{\sc i} D 
($\lambda_{\rm air}^{\rm D2}$=5889.950\AA \, and $\lambda_{\rm air}^{\rm D1}$=5895.924\AA) 
absorption-line spectroscopy (Lehner et al. 1999) towards 12 cluster stars, which 
found that the equivalent widths of the IV component in Ca\,{\sc ii} K range from 0.05 to 0.20 \AA, 
although any variation from star-to-star appeared random and is close to the error in the measurement. 
Langer et al. (1990) also found variations in IVC Na\,{\sc i} D equivalent widths of a factor $\sim$ 3 over 
the cluster face for a handful of stars. Fibre-optic array mapping in the Na\,{\sc i} D absorption 
lines towards the cluster centre (Meyer \& Lauroesch 1999) also found structure visible on scales 
of $\sim$ 4 arcsec at a velocity resolution of $\sim$ 16 km\,s$^{-1}$. Using empirical relationships 
between the Na and H\,{\sc i} column densities, Meyer \& Lauroesch (1999) derived values of the 
H\,{\sc i} column density towards the cluster centre of $\approx$ 5$\times$10$^{20}$ cm$^{-2}$,  
which imply a volume density of $\sim$ 1000 cm$^{-3}$. This compares with the IVC H\,{\sc i} 
column density towards this point, at a resolution of 
$\sim$ 2$\times$1 arcmin, of $\sim$ 4$\times$10$^{19}$ cm$^{-2}$, measured by Smoker et al. (2002) 
using a combined Arecibo-Westerbork Synthesis Radio Telescope map. This H\,{\sc i} map, shown in 
Fig. \ref{combhisur}, demonstrates that the IV H\,{\sc i} column density is actually quite small towards 
the cluster centre, with a stronger IV column density, still relatively near to the cluster 
centre, being at R.A.$\simeq$ 21$^{h}$29$^{m}$40$^{s}$, Dec.$\simeq$ 12$^{\circ}$09$^{\prime}$20$^{\prime\prime}$. 
One of the aims of the current study is thus to determine the Na\,{\sc i} column density 
(and variations over small scales), in areas of the IVC where the H\,{\sc i} is strong. Additionally, 
the H\,{\sc i} observations show indications of multicomponent structure in velocity, perhaps 
indicative of cloudlets. Hence it was decided to search for such structure using higher spectral 
resolution observations than those employed by Meyer \& Lauroesch (1999).

\begin{figure}
\includegraphics[]{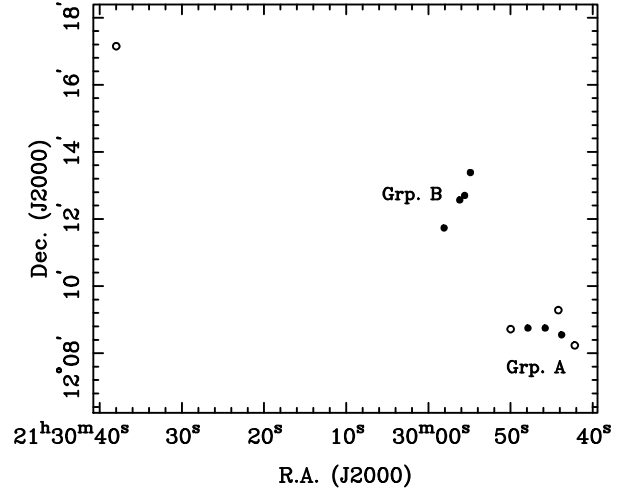}
\caption{Spatial distribution of observed sample. Filled circles are 
cluster stars, open circles are field stars. 
} 
\label{samplefig}
\end{figure}

Candidate stars were chosen from the {\sc simbad} astronomical database where they are all 
listed as being cluster members. However, {\em a-priori} a radial velocity measurement was only 
available for about half of the stars, and, after analysis of the data, it was found that 4 of the 
objects were field stars in the line-of-sight towards M\,15. The majority of the sample are listed 
in Battistini et al. (1985), although also included are the single objects 
KGSY 121 (Kadla et al. 1988), M\,15 ZNG-1 (Zinn et al. 1972), and BRO 24 (Brown 1951). The locations 
of the sample stars are depicted in Fig. \ref{samplefig}. Three cluster stars (group `A' in  
Fig. \ref{samplefig}) lie near a peak in the IVC H\,{\sc i} column density, with 
four cluster stars (group `B') at positions where the IVC H\,{\sc i} column density is 
approximately half of this value. 

%

%
%

Section \ref{obs} describes the observations, Sect. \ref{data} details the data reduction 
and analysis performed to obtain column densities towards the IVC, Sect. \ref{results} gives the 
main results, Sect. \ref{discussion} provides the discussion, and in Sect. \ref{conclusions} 
the conclusions and suggestions for future work are presented.

\section{Observations}
\label{obs}

The observations towards the M\,15 IVC were taken during three sessions in June-August 2001, 
using echelle spectrometers mounted on the Apache Point Observatory 3.5-m (APO3.5), 8.2-m 
Very Large Telescope (UT2; Kueyen), and 4.2-m William Herschel telescope (WHT). The post-AGB 
star whose spectrum was obtained with the VLT, M\,15 ZNG-1, was observed in order to obtain 
its stellar abundance (Mooney et al. 2002), with the interstellar absorption 
components described here being detected serendipitously. In order to determine the contribution 
from the night-sky spectrum to the observed data, 
frequent observations of rapidly-rotating B-type stars were taken which acted as the telluric standards. 
For the APO3.5 and WHT observations, arc line spectra were taken frequently during the night, and hence the 
derived velocities are accurate to better than 0.5 km\,s$^{-1}$, being 3 times the root mean 
square error on the polynomial fit to the arc-line wavelength calibration. For the VLT-observed star, 
arcs were only taken at the start and end of each night, thus the wavelength accuracy depends on the 
stability of the spectrometer. The stellar sample and observing details are listed in Table \ref{obstab}. 

\begin{table*}
\begin{center}
\small
\caption{Observing parameters and stars observed. Names and magnitudes are from 
the {\sc simbad} database and the equinox is J2000. Nomenclature: 
BBFP$nnn$: Battistini et al. (1985); BRO 24: Brown (1951); KGSY 121: 
Kadla et al. (1988); M\,15 ZNG-1: Zinn et al. (1972). Res. refers to the spectral resolution of the 
observations and SNR to the signal-to-noise ratio at a wavelength of $\sim$ 5900\AA. 
$N_{{\rm IVC}}^{\rm HI}$ is the H\,{\sc i} column density measured at a 
resolution of $\sim$2$\times$1 arcmin.
Telescope/instrument names; APO/ARC: Apache Point 3.5-m telescope/Chicago Echelle Spectrograph.
VLT/UVES: Very Large Telescope (Kueyen)/UV-Visual Echelle Spectrograph. 
WHT/UES: William Herschel Telescope/Utrecht Echelle Spectrograph.
}%
\label{obstab}
\begin{tabular}{lrrrrrrrrrr}
\hline
Star      &   R.A.    &   Dec.     & $m_{V}$ & Tel./Instr.&  Obs. Date & Res. &  Time  & SNR & $N_{{\rm IVC}}^{\rm HI}$  & M\,15 \\
          & hhmmss.s  &${\circ}$ \, $^{\prime}$ \, $^{\prime\prime}$ 
                                   & (mag.)  &            & &  (km\,s$^{-1}$) &    (s) &     & ($\times 10^{19}$cm$^{-2}$)&   star? \\
\hline         
BBFP 493   & 212942.2   & +120814  & 14.27  &     APO/ARC  & 9 Jun 2001 & 8.5  &  1400  &  30 &    12  &   No  \\
BBFP 456   & 212943.8   & +120833  & 14.46  &     APO/ARC  & 9 Jun 2001 & 8.5  &  3000  &  45 &    13  &  Yes  \\
BBFP 394   & 212944.2   & +120917  & 13.61  &     APO/ARC  & 9 Jun 2001 & 8.5  &  1400  &  40 &    14  &   No  \\
BBFP 438   & 212945.8   & +120845  & 13.81  &     APO/ARC  & 8 Jun 2001 & 8.5  &  4285  &  45 &    12  &  Yes  \\
BBFP 437   & 212947.9   & +120845  & 13.83  &     APO/ARC  & 9 Jun 2001 & 8.5  &  1800  &  30 &    11  &  Yes  \\
BBFP 442   & 212950.0   & +120843  & 12.66  &     WHT/UES  & 4 Aug 2001 & 6.8  &  6000  &  40 &     8  &   No  \\ 
BBFP 39    & 212954.9   & +121323  & 12.83  &     WHT/UES  & 2 Aug 2001 & 6.2  &  2400  &  45 &     6  &  Yes  \\
BBFP 67    & 212955.6   & +121242  & 13.44  &     WHT/UES  & 2 Aug 2001 & 6.2  &  2400  &  30 &     6  &  Yes  \\
KGSY 121   & 212956.2   & +121234  & 13.02  &     WHT/UES  & 2 Aug 2001 & 6.2  &  2400  &  35 &     6  &  Yes  \\
M\,15 ZNG-1& 212958.1   & +121144  & 14.80  &     VLT/UVES & 3 Jul 2001 & 7.5  & 21600  &  80 &     5  &  Yes  \\
BRO 24     & 213038.0   & +121709  & 13.3   &     WHT/UES  & 2 Aug 2001 & 6.2  &  3600  &  40 &     5  &   No  \\
\hline
\end{tabular}
\normalsize
\end{center}
\end{table*}


\begin{table*}
\begin{center}
\small
\caption{Results for the IVC Na\,D absorption-line measurements. Limiting values for the equivalent widths 
are calculated according to Eq. (\ref{ewlim}). V corresponds to results obtained using Voigt profile fitting,
and D to results obtained using the apparent optical depth method. Upper limits to the column densities 
are 10$\sigma$ limits calculated according to Eq. (\ref{ewlim}) assuming a linear curve of growth. 
}
\label{restab1}
\begin{tabular}{rrrrrcrrrr}
\hline
Star        & $v_{\rm LSR}^{\rm star}$   & $v_{\rm IVC}^{\rm LSR}$(D$_{2}$) &EW$_{\rm star}$(D$_{2}$)&EW$_{\rm IVC}$(D$_{2}$) &log$_{10}N$(D$_{2}$) V, D  &$b$(D$_{2}$) V  & $N_{{\rm NaI}}^{{\rm D2}}/N_{\rm HI}$  \\           
Group       & (km\,s$^{-1}$)             & $v_{\rm IVC}^{\rm LSR}$(D$_{1}$) &EW$_{\rm star}$(D$_{1}$)&EW$_{\rm IVC}$(D$_{1}$) &log$_{10}N$(D$_{1}$) V, D  &$b$(D$_{1}$) V  & $N_{{\rm NaI}}^{{\rm D1}}/N_{\rm HI}$  \\
            &                            & (km\,s$^{-1}$)              &(m\AA)      &(m\AA)             &(log$_{10}$($N$ cm$^{-2}$))                      &(km\,s$^{-1}$)  & ($\times 10^{-8}$)     \\
\hline
BBFP 493    &    $-$24$\pm$3             &      --                     & --         &  $<$50            &\hspace*{-0.85cm}$<$11.48, --      & --                  &   $<$0.3          \\  
(A)         &                            &      --                     & --         &      --           &      --, --                    & --                  &       --          \\
BBFP 456    &  $-$92.6$\pm$0.5           & +68.4$\pm$0.3               & --         & 150$\pm$30        & $>$12.13, $>$12.00             & 3.5$\pm$0.8         &      1.0          \\
(A)         &                            & +68.3$\pm$0.3               & 136$\pm$15 & 150$\pm$30        & $>$12.33, $>$12.26             & 5.4$\pm$0.6         &      1.6          \\
BBFP 394    &  $-$27.5$\pm$1.0           &            --               & --         &  $<$40            &\hspace*{-0.85cm}$<$11.35, --    & --                  &   $<$0.2          \\  
(A)         &                            &            --               & --         &      --           &       --, --                   & --                  &       --          \\
BBFP 438    &  $-$99.0$\pm$0.5           & +66.2$\pm$0.2               & --         & 190$\pm$30        & $>$12.30, $>$12.25             & 5.6$\pm$0.3         &      1.7          \\  
(A)         &                            & +65.9$\pm$0.1               & 193$\pm$18 & 200$\pm$20        & $>$12.55, $>$12.50             & 5.5$\pm$0.3         &      3.0          \\
BBFP 437    & $-$102.3$\pm$0.5           & +68.5$\pm$0.4               & --         & 200$\pm$20        & $>$12.27, $>$12.24             & 5.7$\pm$0.5         &      1.7          \\
(A)         &                            & +68.1$\pm$0.4               & 213$\pm$20 & 190$\pm$30        & $>$12.56, $>$12.47             & 5.3$\pm$0.6         &      3.3          \\
BBFP 442    &   +19$\pm$5                &            --               & --         & $<$35             &\hspace*{-0.85cm}$<$11.33, --    & --                  &   $<$0.3          \\  
(A)         &                            &            --               & --         &      --           &      --, --                    & --                  &       --          \\
BBFP 39     &$-$100.0$\pm$0.5            & +63.7$\pm$0.4               & 251$\pm$12 &  97$\pm$6         & 12.00$\pm$0.03, 12.02$\pm$0.03 & 4.8$\pm$0.3         &      1.7          \\
(B)         &                            & +64.3$\pm$0.3               & 230$\pm$12 &  93$\pm$9         & 12.06$\pm$0.04, 12.09$\pm$0.04 & 5.3$\pm$0.4         &      1.9          \\
BBFP 67     &$-$101.0$\pm$0.5            & +63.0$\pm$0.3               & 226$\pm$14 & 131$\pm$12        & 12.02$\pm$0.04, 11.93$\pm$0.04 & 3.6$\pm$0.2         &      1.7          \\
(B)         &                            & +63.4$\pm$0.4               & 222$\pm$14 &  76$\pm$8         & 11.98$\pm$0.05, 11.94$\pm$0.05 & 4.0$\pm$0.4         &      1.6          \\
KGSY 121    & $-$96.5$\pm$0.5            & +61.0$\pm$0.3               & 256$\pm$13 & 129$\pm$10        & 11.99$\pm$0.03, 12.00$\pm$0.03 & 4.7$\pm$0.4         &      1.6          \\
(B)         &                            & +60.9$\pm$0.3               & 245$\pm$15 &  79$\pm$6         & 12.06$\pm$0.03, 12.03$\pm$0.03 & 3.7$\pm$0.4         &      1.9          \\
M\,15 ZNG-1 &$-$100.0$\pm$1.0            & +64.4$\pm$0.2               &  --        &  87$\pm$5         & 11.80$\pm$0.04, 11.79$\pm$0.04 & 3.8$\pm$0.4         &      1.3          \\ 
(B)         &                            & +63.5$\pm$0.3               &  21$\pm$4  &  43$\pm$4         & 11.79$\pm$0.04, 11.79$\pm$0.04 & 3.7$\pm$0.4         &      1.2          \\  
BRO 24      &   +22$\pm$5                &           --                &  --        &  $<$35            &\hspace*{-0.85cm}$<$11.33, --   & --                  &   $<$0.4          \\  
--          &                            &           --                &  --        &      --           &      --, --                    & --                  &       --          \\
\hline
\end{tabular}
\normalsize
\end{center}
\end{table*}


\begin{table*}
\begin{center}
\small
\caption{Results for the IVC Ca\,{\sc i} (4226\AA) and Ca\,{\sc ii} K (3933\AA) absorption-line measurements 
towards M\,15 ZNG-1. For the Ca\,{\sc ii} K-line, a two-component Voigt fit was performed to the data. V 
corresponds to results obtained using Voigt profile fitting, and D to results obtained using the apparent 
optical depth method. Upper limits to the Ca\,{\sc i} column densities are 10$\sigma$ limits calculated according to 
Eq. (\ref{ewlim}) assuming a linear curve of growth. 
}
\label{restab2}
\begin{tabular}{rrrrrrr}
\hline 
 Transition  & $v_{\rm LSR}^{\rm star}$ & $v_{\rm IVC}^{\rm LSR}$      & EW            &(log$_{10}$($N$ cm$^{-2}$)) V, D & $b$  V         \\  
             & (km\,s$^{-1}$)           & (km\,s$^{-1}$)               & (m\AA)        &                                 & (km\,s$^{-1}$) \\  
\hline
Ca\,{\sc ii} K (3933\AA)  & $-$100.0$\pm$1.0 & 52.7$\pm$0.6 &   6$\pm$1  & 11.16$\pm$0.07, --                            & 0.6$\pm$4.0    \\
   $\prime\prime$         & $\prime\prime$   & 64.9$\pm$0.1 & 122$\pm$4  & 12.40$\pm$0.02, 12.36$\pm$0.02                & 5.3$\pm$0.2    \\
Ca\,{\sc i} (4226\AA)     & $\prime\prime$   & --           & $<$20      &    $<$10.9, --                                & --             \\
\hline
\end{tabular}
\normalsize
\end{center}      
\end{table*}

%
%

\section{Data reduction and analysis}
\label{data}

The APO3.5 and WHT data were reduced using standard methods within {\sc iraf}{\it \footnote{ {\sc iraf} 
is distributed by the National Optical Astronomy Observatories, U.S.A.}}. This included debiasing, wavelength 
calibration and extraction of the individual orders using {\sc doecslit}. No flatfielding was performed. 
For the APO3.5 observations, the limited height of the slit (3.0 arcsec) precluded sky subtraction, and hence 
we performed a single 750-s exposure on the dark sky to estimate the contribution of the full moon to 
the night sky brightness. This added between 10$-$20 per cent to the continuum level for our targets, 
and was subtracted before the extraction was performed. 

For the VLT-observed star, we used data products obtained from the pipeline-calibration which includes 
wavelength calibration, sky subtraction and flatfielding. 
The extracted spectra were imported into {\sc dipso} (Howarth et al. 1996), shifted to the Dynamical Local 
Standard of Rest (LSR) using values obtained from the {\sc starlink} programme {\sc rv} 
(Wallace \& Clayton 1996), and normalised by fitting low-order polynomials to the continuum. 
The radial velocities of the stars were determined by use of the two stellar Na\,D lines for the cluster objects, 
and where possible using Ca\,{\sc i} ($\lambda_{\rm air}$=6717.681\AA) for the field stars, with the exception of 
M\,15 ZNG-1 for which we used the Si\,{\sc iii} features ($\lambda_{\rm air}$=4552.622\AA, 4567.840\AA). The 
broader H$\delta$ and H$\beta$ lines were used as a check on the derived velocity. 
Finally, the normalised spectra were analysed using both the apparent optical depth (AOD) method  
(Savage \& Sembach 1991) and standard 
Voigt profile fitting within {\sc dipso} using the {\sc elf} and {\sc is} programs, for which we used oscillator 
strengths of 0.6340, 0.318, 0.6306 and 0.33919 for the Ca\,K, Na\,D$_{1}$, Na\,D$_{2}$ and K\,{\sc i} 
($\lambda_{\rm air}$=7698.974\AA) lines, respectively (Welty et al. 1999). 
The derived fits were convolved with an instrumental $b$-value of 0.6006 times the FWHM shown in 
Table \ref{obstab}. The errors on the derived equivalent widths (EWs) were estimated from a combination of the baseline uncertainty 
plus the photon noise, and those on the column densities from the EW error, assuming a linear curve 
of growth. 

\section{Results}
\label{results}

Figures \ref{NaDspectra} and \ref{m15zng1} show the Na\,D spectra towards the 11 sample stars. 
Inspection of the telluric spectra indicate that the only possible contaminating features caused by 
atmospheric absorption are at $\sim$ 5896.46\AA \, and $\sim$ 5893.83\AA, which have equivalent 
widths of $\sim$ 5 m\AA, compared to $\sim$ 90$-$200 m\AA 
\, for the IV Na\,D$_{2}$ equivalent widths. Additionally, for both the APO3.5 
and WHT spectra, there is some narrow emission present at the wavelengths of the Na\,D doublet 
(5889.950\AA \, and 5895.924\AA), presumably caused by street lights. This ranged in LSR 
velocity from $\sim$+20 to +35 km\,s$^{-1}$, so was just far enough removed in velocity so as not to 
contaminate the IV absorption-line at $\sim$ 65 km\,s$^{-1}$. 

The spectra are composite, and in 7 cases show a stellar component at $\approx$ $-$100 km\,s$^{-1}$, 
low-velocity (LV) components centred upon $\sim$ 0 km\,s$^{-1}$ and intermediate-velocity (IV) 
components at $\sim$ +60$-$70 km\,s$^{-1}$. Velocities of the M\,15 cluster stars range from 
$\sim$ $-$93 to $-$102 km\,s$^{-1}$, and, at their radial distance from the centre, are each within the 
range of velocities observed towards M\,15 by Drukier et al. (1998): the cluster having a mean LSR 
velocity of $-$97.1 km\,s$^{-1}$ and velocity dispersion that changes with distance from 
the centre. For the M\,15 cluster stars (excepting M\,15 ZNG-1, which is a pAGB), we measure 
Na\,{\sc i} D$_{1}$ equivalent widths of between 140 and 250 m\AA, corresponding to A$-$F spectral 
types (Gratton \& Sneden 1987; Jaschek \& Jaschek 1987).

Tables \ref{restab1} and \ref{restab2} show the results of the profile-fitting
to the IV interstellar spectra for the Na\,D and Ca\,K lines, respectively. This yields the central 
velocities, $b$-values and column densities of the IV components. For the three cluster stars in 
group `A' (near a peak in the IVC H\,{\sc i} column density), the derived column densities in 
the Na\,D$_{1}$ line are $\sim$ 0.25 dex higher than those obtained for the Na\,D$_{2}$ line, indicating 
that saturation is occurring in these sightlines, and that the derived column densities are 
lower limits. 

Towards the non-detections, we estimated a 10$\sigma$ limiting equivalent width that would be 
detected using the current data, EW$_{\rm lim}$, thus;

\begin{equation}
{\rm EW_{\rm lim}}=10\sigma_{\rm cont} \Delta\lambda_{\rm instr},
\label{ewlim}
\end{equation}

\noindent
where $\sigma_{\rm cont}$ is 1/(continuum signal-to-noise ratio) and $\Delta\lambda_{\rm instr}$ 
is the instrumental full width half maximum (FWHM) in \AA. The limiting values of the Na\,{\sc i} 
column densities were estimated by using the EW$_{\rm lim}$ values estimated above, and $b$-values 
corresponding to the nearest star for which there were an IV gas detection, assuming that we are 
on the linear part of the curve of growth. Towards M\,15 ZNG-1, the above equation was also used to 
derive upper limits to the column densities of Fe\,{\sc i} ($\lambda_{\rm air}$=3859.9111\AA), 
Al\,{\sc i} ($\lambda_{\rm air}$=3944.0060\AA) and K\,{\sc i} ($\lambda_{\rm air}$=4044.143\AA). 
The 10$\sigma$ upper limits of log$_{10}$($N$ {\rm cm$^{-2}$) of these species of 12.6, 11.9 and 13.1 
are high, and are not discussed further in this paper. 

Figure \ref{KIspectra} shows the K\,{\sc i} spectra of the five stars observed with the APO3.5 
(the other two observing runs not covering this wavelength region). In three 
of the sightlines, there are indications of IV absorption in K\,{\sc i}. Although the signal-to-noise
of the data are low, we believe that the intermediate-velocity absorption detected towards the 
cluster stars is likely to be interstellar in nature because; (1) There are no telluric 
features at the same wavelength as the IV absorption, either in the two field stars or 
telluric standards; (2) The features are unlikely to be stellar because the centres of two 
nearby lines (blueshifted by the stellar velocity) of Nb\,{\sc i} ($\lambda_{\rm air}$=7703.33\AA) 
and Cl\,{\sc i} ($\lambda_{\rm air}$=7702.829\AA) would be more than 4 km\,s$^{-1}$ away from the 
IV absorption peak (at least for the stars BBFP 438 and BBFP 456); (3) The FWHM velocity widths of the 
IV absorption of $\sim$ 7$-$10 km\,s$^{-1}$ (uncorrected for instrumental broadening: the 
components are unresolved) are somewhat narrower than the Na\,D$_{1}$ stellar profiles which 
have uncorrected FWHM velocity widths from 13$-$16 km\,s$^{-1}$. As a species, K\,{\sc i} 
tends to have narrow velocity components, the median width FWHM of 54 Galactic sightlines 
observed by Welty \& Hobbs (2001) being $\sim$ 1.2 km\,s$^{-1}$. Equivalent widths, velocities, 
and column densities of the IV K\,{\sc i} components are shown in Table \ref{restab3}. 


%
\begin{figure*}
\includegraphics[]{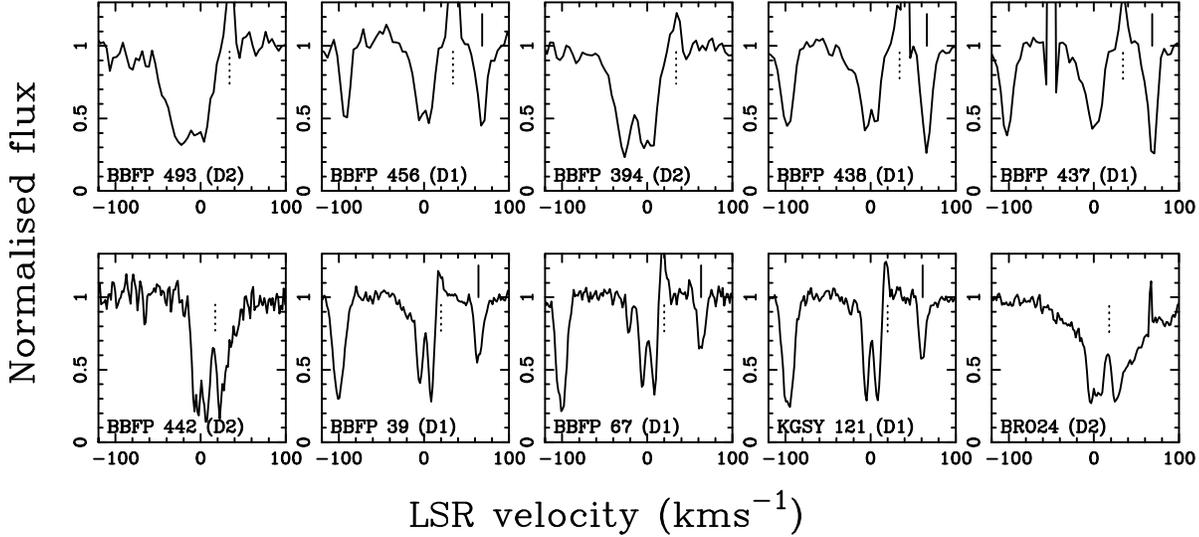}
\caption{Na\,D absorption-line spectra for the whole sample apart from M\,15 ZNG-1 (shown in 
Fig. \ref{m15zng1}). For IV detections the less-saturated Na\,D$_{1}$ line is shown, for 
non-detections the more sensitive Na\,D$_{2}$ line is depicted.
The abscissa shows the LSR velocity and the ordinate the normalised 
flux. Solid vertical lines indicate the velocity centroids of the IVC absorption. Dashed 
vertical lines indicate the presence of emission caused by street lights.
} 
\label{NaDspectra}
\end{figure*}

\begin{figure}
\includegraphics[]{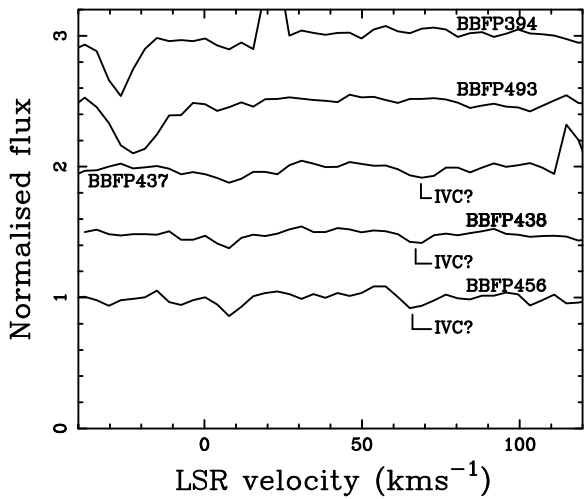} 
\caption{K\,{\sc i} ($\lambda_{\rm air}$=7698.974\AA) absorption-line spectra for 
five stars observed with the APO3.5. The top two objects are field stars, with the 
bottom three being cluster stars. The abscissa shows the LSR velocity and the ordinate 
the normalised flux, offset in units of 0.5 for clarity. 
} 
\label{KIspectra}
\end{figure}

\begin{table}
\begin{center}
\small
\caption{Results for the tentative detection of IV gas in K\,{\sc i} ($\lambda_{\rm air}$=7698.974\AA) 
towards three M\,15 cluster stars. Voigt fitting was used to determine the column  
densities.
}
\label{restab3}
\begin{tabular}{lrrr}
\hline 
                                                                 &   BBFP 437     &  BBFP 438    &  BBFP 456       \\
$v_{\rm IVC}$ (km\,s$^{-1}$)                                     &  68.7$\pm$1.0  & 67.0$\pm$0.5 &  66.6$\pm$1.0   \\
EW$_{\rm IVC}$ (m\AA)                                            &    28$\pm$8    &   18$\pm$5   &   24$\pm$8      \\
log$_{10}$($N_{\rm IVC}^{\rm KI} \, {\rm cm^{-2}}$)                                           & 11.20$\pm$0.30 &11.10$\pm$0.30&  11.15$\pm$0.30 \\ 
$N_{\rm KI}/N_{\rm HI}$ ($\times$10$^{-9}$)  & 1.4            & 1.0          &  1.1            \\
\hline
\end{tabular}
\normalsize
\end{center}
\end{table}

\section{Discussion}
\label{discussion}

\subsection{The distance to the M\,15 IVC} 

We note that in principle the fact that we have obtained non-detections towards 4 field stars 
in Na\,D could enable us to determine a lower-distance limit to this part of the IVC, {\em if} the distances to 
the objects were known. 
Of the three field stars, two (BBFP 394 and BBFP 493), have $V$ magnitudes 
in the range delineated by the cluster stars in the current paper (see Buonanno et al. 1983), 
with BBFP 442 being towards the bright end of the cluster star magnitude range 
($m_{V}$=12.46). The ($B-V)$ and $(U-B)$ colours of the field stars are $\sim$ 0.2$-$0.3 mag and 0.2$-$0.6 mag. bluer, 
respectively, than the median values for our sample of cluster stars. With the current data, due to the
difficulty in ascribing a luminosity class for the field stars, we have not estimated their distances. 
Additionally, 
we note that a recent compilation of IVC and HVC Na\,D column densities by Wakker (2001) 
has indicated that to be sure of detecting an interstellar absorption in Na\,D towards an IVC with 
$N_{\rm HI}$=10$^{19}$ cm$^{-2}$, a signal-to-noise ratio (SNR) exceeding $\sim$ 1000$-$2000 is required. 
Hence, even though the IV H\,{\sc i} column densities exceed 10$^{20}$ cm$^{-2}$ 
towards three of our field-star sightlines, the observed SNRs of $\sim$ 30$-$40 in the stars without an IVC 
detection preclude us from making any statements about the distance to this part of 
the IVC. 

%

\subsection{Two-component velocity structure}

The presence of two-component velocity structure in IVCs and HVCs indicates the 
presence of cloudlets, collisions between which have been postulated as a mechanism for 
star formation in the halo of the Galaxy (Dyson \& Hartquist 1983, although see 
Christodoulou et al. 1997 and Ramspeck et al. 2001). One of our observed stars, BRO\,24, was chosen 
as a target as it lies near to a position in the IVC H\,{\sc i} map where there are 
indications of two-component structure (Smoker et al. 2002). Unfortunately, this 
star turned out to be a field star. However, towards the object M\,15 ZNG-1, we 
did detect two components (at $\sim$ +53 and +64 km\,s$^{-1}$), in both 
Ca\,K and Na\,D, depicted in Fig. \ref{m15zng1}, indicating overlapping velocity 
components in the line-of-sight. 

\begin{figure}
\includegraphics[]{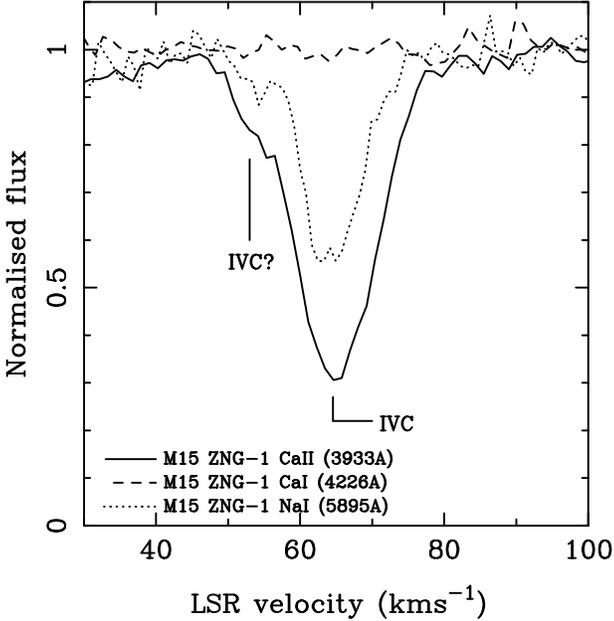}
\caption{IVC absorption towards the globular cluster star M\,15 ZNG-1, showing possible 
two-component velocity structure in both Ca\,K and Na\,D$_{1}$ at $\sim$ +53 and +64 km\,s$^{-1}$. 
No IVC absorption is evident from the Ca\,{\sc i} line at $\lambda_{\rm air}$=4226.728\AA.
Absorption bluewards of $\sim$ +40 km\,s$^{-1}$ originates from the wing of the LV gas. 
} 
\label{m15zng1}
\end{figure}

%
%

\subsection{Na\,{\sc i} and H\,{\sc i} column densities and small-scale variations}

Wakker (2001) has summarized the presently-existing absorption-line data for all species 
reported towards H\,{\sc i} clouds, finding that, for LV gas, there is a mean Na\,{\sc i}/H\,{\sc i} column 
density ratio of $\sim$ 5$\times$10$^{-9}$, but with a large range exceeding a factor of 10. 
Some IVCs (such as the LLIV arch) have similar values to this mean, with others, such 
as previous measurements towards the M\,{\sc 15} IVC (Complex gp) showing 
somewhat higher values, ranging from (10$-$160)$\times$10$^{-9}$ (Langer et al. 1990; 
Kennedy et al. 1998b; Meyer \& Lauroesch 1999). The Na\,{\sc i}/H\,{\sc i} column density ratios 
for the current dataset (based upon H\,{\sc i} data of resolution $\sim$ 1$\times$2 arcmin), 
vary from $\sim$ (1$-$3)$\times$10$^{-8}$, and by as much as a factor of $\sim$ 2 in sightlines 
only separated by 30 arcsec. 

Towards the 7 M\,15 cluster stars, we derive logarithmic column densities ($N$ in cm$^{-2}$), 
log$_{10}$($N$), of between $\sim$ 11.8$-$12.5, the latter value being a lower limit due 
to saturation. This corresponds to a factor $\sim$ 6 variation in column density 
between parts of group `A' and `B', which compares to a variation in Na\,{\sc i} IVC column 
density of $\sim$ 16 over less than an arcminute towards the centre of the cloud 
(Meyer \& Lauroesch 1999).

For the `A' group of 3 stars, within the errors, two (BBFP 437, 438)
have the same value of Na\,{\sc i} column density, with BBFP 456 (separated by 
$\sim$ 30$^{\prime\prime}$ from BBFP 438), having a value $\sim$ 70 per cent lower 
than the other two. A rough estimate of the volume density can be made using 
these data. Assuming that the IVC is at a distance of 1 kpc, the separation between 
BBFP 456 and the other two stars is $\sim$ 0.15 pc and the approximate transverse cloud size at 
full width half maximum column density is thus $\sim$ 0.3 pc. Following Shaw et al. (1996) 
and Kennedy et al. (1998b), {\em if} this part of the cloud is spherically symmetric, 
and assuming a $N$(Na\,{\sc i})/$N$(H\,{\sc i}) ratio of 4$\times$10$^{-9}$ appropriate to 
high-latitude clouds (Ferlet et al. 1985), the resulting volume density for 
log$_{10}(N$(Na\,{\sc i} cm$^{-2}$))=12.55 is $n_{\sc HI}$ $\approx$ 1000 cm$^{-3}$. 
Clearly if there is sheet-like structure in the IVC, this volume density is an upper limit, 
and hence we emphasize that given the difficulty in determining the $N$(Na\,{\sc i})/$N$(H\,{\sc i}) 
ratio and unknown cloud geometry, this implied volume density is very uncertain.

Within the `B' group, three of the stars display similar 
Na\,{\sc i} column density values, apart from M\,15 ZNG-1, which is deficient by 
$\sim$ 50 per cent, being separated by $\sim$ 1 arcmin from its closest observed companion, KGSY 121. 
Towards the stars ARP IV-38 and ARP II-75 (Arp 1955), Kennedy et al. (1998b) found $N$(Na\,{\sc i}) 
values of $\sim$ 1.6$\times$10$^{12}$ cm$^{-2}$ and 2.5$\times$10$^{13}$ cm$^{-2}$. 
%
%
Using our IVC H\,{\sc i} data of resolution 
$\sim$ 2$\times$1 arcmin, these correspond to  $N$(Na\,{\sc i})/$N$(H\,{\sc i}) values of 
4$\times$10$^{-8}$ and 2.5$\times$10$^{-7}$, respectively. The latter value is some $\sim$ 25 
times greater than observed towards M\,15 ZNG-1 and is typical of large variations (often exceeding 
a factor of 10) that can occur in IVCs/HVCs (Wakker 2001). Possible reasons for this are discussed in 
Meyer \& Lauroesch (1999). 

\subsection{Cloud conditions towards the M\,15 ZNG-1 sightline}

Towards the PAGB star M\,15 ZNG-1, the H\,{\sc i} column density observed at
a resolution of $\sim$ 2$\times$1 arcmin is $N_{\rm HI}$=5$\times$10$^{19}$ cm$^{-2}$. 
Our echelle data of SNR $\sim$ 80 cover Ca\,{\sc i} ($\lambda_{\rm air}$=4226.728\AA), Ca\,{\sc ii} K 
and the Na\,{\sc i} D lines, thus providing some information on the properties of the IV gas by use 
of these unsaturated features. We determine logarithmic column density values for Ca\,{\sc ii}, Ca\,{\sc i} 
and Na\,{\sc i} of $\sim$ 12.4, $<$10.9 and 11.8, respectively. The Ca\,{\sc i}/Ca\,{\sc ii} column 
density ratio is $<$0.03. We note that this value is consistent with the composite standard model of 
the photoionised warm interstellar medium of Sembach et al. (2000) (their Table 5), that gives the 
ionisation fraction Ca\,{\sc i}/Ca\,{\sc ii} $\sim$ 0.02. 

Our derived value of $N$(Ca\,{\sc ii})/$N$(H\,{\sc i}) $\sim$ 5$\times$10$^{-8}$ for 
the IVC is within the {\em range} found for the high Galactic latitude clouds 
studied by Albert et al. (1993), namely (0.3--30)$\times$10$^{-8}$ cm$^{-2}$. 
However, using the observed H\,{\sc i} column density in combination with 
the correlation plot of the Ca\,{\sc ii} abundance against H\,{\sc i} column density for 
IVCs/HVCs from Wakker \& Mathis (2000), a predicted Ca\,{\sc ii} column density 
of $\sim$ 1$\times$10$^{-8}$ cm$^{-2}$ is obtained. This is $\sim$ 5 times smaller than the observed value.
Our derived value of $N$(Ca\,{\sc ii})/$N$(H\,{\sc i}) $\sim$ 5$\times$10$^{-8}$ implies 
a depletion of $\sim -$1.6, given a Solar calcium abundance of 2.10$\times$10$^{-6}$ and 
{\em assuming} that Ca\,{\sc ii} is the dominant ionisation species. However, the model of 
the photoionised warm interstellar medium of Sembach et al. (2000), 
predicts that the Ca\,{\sc ii}/Ca\,{\sc iii} fraction is $\sim$ 0.26. This implies 
that the Ca\,{\sc iii} ion is likely to be dominant, particularly as the detection 
of H$\alpha$ emission from the cloud (Smoker et al. 2002) indicates that the gas is partially 
ionised (although perhaps by collisions and not photoionisation; Sembach 1995). 

The $N$(Na\,{\sc i})/$N$(H\,{\sc i}) ratio of $\sim$ 1.3$\times$10$^{-8}$ towards this sightline 
compares with the value derived using the correlation for IVCs/HVCs from Wakker \& Mathis (2000) of 
$\sim$ 0.7$\times$10$^{-8}$. Combined with the Ca result, this again indicates that 
this part of Complex gp has high ratios of these elements with respect to H\,{\sc i} when 
compared with other HVCs and IVCs (c.f. Wakker 2001). Part of this `overabundance' may be 
due to the fact that the cloud is partially ionised: Smoker et al. (2002) find that the M\,15 IVC has an 
H$\alpha$ flux exceeding 1 Rayleigh, which, using a reasonable value for the size of 
the cloud and the calculated emission measure, produces a H\,{\sc ii}/H\,{\sc i} ratio 
of $\sim$ 1$-$2$\times f^{1/2}$, where $f$ is the filling factor of H\,{\sc ii} in the IVC. 

Towards this sightline, the IVC Na\,{\sc i}/Ca\,{\sc ii} column density ratio is $\sim$ 0.25$\pm$0.03. 
Within the interstellar medium, the range of observed values is from $\sim$ 0.1$-$100; 
high ratios being found in dense, cold clouds where Ca is depleted onto dust, and 
low values occurring in the local warm interstellar medium, where heating of dust grains, 
for example by shocks, causes grain destruction and a mean Na\,{\sc i}/Ca\,{\sc ii} ratio 
of $\sim$ 0.22 (Bertin et al. 1993). In the local ISM, some 15$-$25 per cent of H\,{\sc i} 
clouds with anomalous velocities exceeding 20 km\,s$^{-1}$ show such low ratios (see Bertin et 
al. 1993 and references therein). 
Grain destruction by shocks is consistent with the observation 
of strong H$\alpha$ emission, perhaps caused by collisional ionisation. However, grain 
destruction does not appear to be total, as Smoker et al. (2002) showed that the 60$\mu$ IRAS 
map and IVC H\,{\sc i} column density map show spatial correspondence, tentatively indicating the 
presence of some dust in the M\,15 IVC. Additionally, large variations in the Na\,{\sc i}/Ca\,{\sc ii} 
ratio have been found to occur in components separated by only a few km\,s$^{-1}$, and 
hence Welty et al. (1996) caution against using this ratio as an indicator of either cloud 
physical conditions or calcium depletions. 

\subsection{Tentative K\,{\sc i} detections}

There are few observations of K\,{\sc i} ($\lambda_{\rm air}$=7698.974\AA) 
in IVCs and HVCs (Wakker 2001). Because of its relatively low abundance compared 
with Ca or Na, combined with a low ionisation potential, K\,{\sc i} is normally used 
to probe relatively dense and cool areas of the interstellar medium. Radio-line H\,{\sc i} 
observations towards the peaks of the M\,15 IVC imply upper limits to the kinetic temperature 
(in the absence of turbulence and averaged over $\sim$ 1 arcmin squared) of less than $\sim$ 500 K 
(Smoker et al. 2002), and H\,{\sc i} column densities exceeding 10$^{20}$ cm$^{-2}$. Hence {\em a priori} 
there was some hope that K\,{\sc i} would be detected, given the fact that Na and Ca are 
relatively abundant in the cloud. 

The only previously-known detections of K\,{\sc i} in IVCs are towards SN 1993J (Vladilo et al. 1994) and 
towards the M\,15 IVC by Kennedy et al. (1998b). The latter found log$_{10}$($N$(K\,{\sc i}) cm$^{-2}$)
$\sim$ 11.4 for the cluster star M\,15 ARP II-75. Using our high-resolution H\,{\sc i} map, 
which has $N_{\rm IVC}$(H\,{\sc i}) $\sim$1$\times$10$^{20}$ cm$^{-2}$, this corresponds 
to $N$(K\,{\sc i})/$N$(H\,{\sc i}) $\sim$ 2.5$\times$10$^{-9}$. Towards another cluster star, 
M\,15 ARP II-38, towards which our H\,{\sc i} IVC column density map has $N_{\rm IVC}$(H\,{\sc i}) 
$\sim$ 4$\times$10$^{19}$ cm$^{-2}$, Kennedy et al. (1998b) obtained an upper limit to 
log$_{10}$($N$(K\,{\sc i}) cm$^{-2}$) $\sim$ 10.6, or 
$N$(K\,{\sc i})/$N$(H\,{\sc i}) $<$ 1$\times$10$^{-9}$. 

These results are similar to our K\,{\sc i}/H\,{\sc i} column density ratios of $\sim$ 0.5$-$3$\times$10$^{-9}$ 
(taking into account the large uncertainty in the measurements) for three closely-spaced stars. 
Based on the uncertainty in our IVC K\,{\sc i} detections, we choose not to over-interpret our 
results. However, we note that both the current data, and that of Kennedy et al. (1998b), 
display $N$(K\,{\sc i})/$N$(H\,{\sc i}) values more than 10 times greater than $N$(K\,{\sc i})/$N$(H$_{\rm tot}$) 
ratios (where $N$(H$_{\rm tot}$)=H\,{\sc i} + 2\,H$_{2}$) observed towards Galactic sightlines 
(Fig. 17 of Welty \& Hobbs 2001). This may be due to a combination of different halo/disc depletions 
for K, ionisation of the H\,{\sc i} in the IVC (and possible H$_{2}$ content?), and the fact that 
our H\,{\sc i} value is derived from data of poorer spatial resolution than the K\,{\sc i} data. Finally, 
we note that the N(K\,{\sc i})/N(Na\,{\sc i}) values are less than 3 times as large in the IVC as 
compared with the Galactic results of Welty \& Hobbs (2001), indicating that the high 
$N$(K\,{\sc i})/$N$(H\,{\sc i}) value may be caused by one of the factors listed above. 

\section{Summary and Conclusions}
\label{conclusions}

At a resolution of $\sim$ 6.2$-$8.5 km\,s$^{-1}$, we have observed absorption in Na\,{\sc i} D at 
intermediate velocities towards 7 stars within the globular cluster M\,15. One of these sightlines was  
also observed in Ca\,{\sc ii} K, and three in K\,{\sc i}. 
The Na\,{\sc i} and Ca\,{\sc ii} to H\,{\sc i} column density ratios 
lie within the range previously observed in high latitude clouds, although both are in the upper echelons 
of the observed distribution for IVCs and HVCs, perhaps partly caused by ionisation of H. Variations in 
the Na\,{\sc i} column density of between 0 to $\sim$ 70 per cent towards the IVC have been detected on 
scales of $\sim$ 30 arcsec (or $\sim$ 0.15pc, assuming a cloud distance of 1 kpc). Over the whole 
cloud, the Na\,{\sc i}/H\,{\sc i} column density ratio varies by as much as a factor of $\sim$ 25. 

The Na\,{\sc i}/Ca\,{\sc ii} column density ratio of $\sim$ 0.25 is typical of gas in the local warm 
interstellar medium. This, and the presence of strong H$\alpha$ emission towards this sightline, 
raises the possibility that shock ionisation and grain destruction is occurring in the IVC, although 
the presence of weak IRAS flux implies that grain destruction is not total. Three cluster stars also 
show tentative detections of K\,{\sc i} (7698.974\AA), with $N$(K\,{\sc i})/$N$(H\,{\sc i}) ratios of 
$\sim$ 0.5$-$3$\times$10$^{-9}$.

Future observations should concentrate on determining the distance to this part of IVC Complex gp, 
via spectroscopy of field stars located towards regions of strong IVC H\,{\sc i}  
column density. Towards the H\,{\sc i} peaks, a signal-to-noise exceeding $\sim$ 20 in Ca\,K 
would be sufficient at a 10$\sigma$ level to interpret a non-detection as being a lower-distance 
limit, allowing for a factor of 5 variation in the $N$(Ca\,{\sc ii})/$N$(H\,{\sc i}) ratio and 
with a resolution of $\sim$ 0.1\AA. Finally, UV observations towards the IVC would much better 
constrain the gas parameters in the IVC. Although the faintness of the M\,15 stars precludes 
this with currently-existing space-based facilities, progress may be possible using the planned 
Cosmic Origins Spectrograph on the Hubble Space Telescope.

%
%

\begin{acknowledgements}

Based on observations obtained with the Apache Point Observatory 3.5-meter telescope, 
which is owned and operated by the Astrophysical Research Consortium, U.S.A., 
the William Herschel Telescope, operated by the Isaac Newton Group of Telescopes,
La Palma, Spain, and at the European Southern Observatory, Cerro Paranal, Chile 
(programme ID 67.D-0010A). We would like to thank an anonymous referee for 
useful comments. JVS is grateful to {\sc pparc} for financial support and 
{\sc starlink} for providing computer facilities. 

\end{acknowledgements}

{}
\end{document}